\documentclass[aps,pre,twocolumn,10pt,groupedaddress,showpacs]{revtex4-1}
\usepackage{graphicx}
\usepackage{notes2bib}
\usepackage{siunitx}
\usepackage{color}
\usepackage{textcomp}
\usepackage{amsmath}
\usepackage{hyphenat}
\newcommand*{\redbullet}{\textcolor{red}{\textbullet}}
\begin{document}

%\preprint{}

\title{Towards attosecond pulse synthesis from solids: spectral shaping, field autocorrelation, and two-color harmonic generation}

\author{Andre Seyen}
\author{Roland Kohn}
\author{Uwe Bovensiepen}
\author{Dietrich von der Linde}
\author{Alexander Tarasevitch}
\email{alexander.tarasevitch@uni-due.de}
%\thanks{}
\affiliation{Fakult\"{a}t f\"{u}r Physik, Universit\"{a}t Duisburg-Essen, Lotharstra{\ss}e 1, Duisburg 47057, Germany}

\date{\today}

\begin{abstract}
Harmonic generation from solid surfaces is a promising tool for producing high energy attosecond pulses. We report shaping of the harmonic spectrum to achieve the bandwidth necessary for attosecond pulse generation. The shaping is demonstrated for lower as well as for higher harmonics using single and two-pulse pumping. The measured harmonic field autocorrelation function exhibits attosecond spikes in good agreement with the harmonic spectrum. Double slit experiments reveal a high spatial coherence of the harmonic beam.
\end{abstract}

\maketitle

\section{\label{intro}
Introduction}
The investigation of the fundamental steps of electron dynamics requires attosecond time resolution~\cite{corkum:attsc,krausz:ivanov,goul:nat10}. A significant development in this field became possible due to the success in attosecond pulse production using high order harmonic generation (HOHG) in gases~\cite{L'Huillier:hhg,gouliemakis:single}. Although XUV radiation with the energy per pulse exceeding 10~$\mu J$ has been demonstrated~\cite{Takahashi}, its level is limited by the onset of ionization in the gas. Much higher energies can be expected using HOHG from solid surfaces~\cite{vdl:jenke,mourou:relopt,tsakiris:atto,naumova:atto,pukhov:extreme,baeva:theory,tar:JPB}. In this case the harmonics are generated using femtosecond pump pulses with intensities $I_p> 10^{17}$~$W/cm^2$. At such intensities the target surface is highly ionized by the leading edge of the pump pulse, and a reflecting layer of overcritical plasma is formed. Such a plasma layer can withstand very high laser intensities extending well into the regime when the motion of the plasma electrons becomes relativistic. Consequently, much higher nonlinearities may be induced.

The mechanism of the HOHG from surfaces can be qualitatively explained using the simple physical model of an oscillating plasma mirror~\cite{tar:JPB,bulanov:mirr,lichters:PP96,gibbon:pic,vdl:rzh,vdl:apb,tar:PRL07}. Due to the femtosecond duration of the pump pulses there is no time for significant plasma expansion. This leads to formation of a steep gradient of the plasma density with a scale length $L\ll\lambda$, where $\lambda$ is the pump wavelength. If the details of the electron density distribution are neglected, the collective electron motion created by the incident electromagnetic wave can be considered as an oscillating mirror.

At intensities $I_p<10^{18}$~$W/cm^2$ the harmonics are produced by the ``classical'' mirror nonlinearity due to electron motion through the steep boundary~\cite{harmCO2:Bezzer}. In this case the harmonic efficiency is strongly dependent on the density gradient. Under certain conditions the harmonic emission can be much enhanced by a resonant excitation of local plasma oscillations~\cite{tar:PRL07,lichters:97,quere:PRL06,tar:EJP,dromey:PRL09}, which is often addressed as CWE (coherent wake emission) regime. At higher intensities the relativistic nonlinearity comes into play~\cite{tsakiris:atto,tar:JPB,tar:PRL07,pukhov:zepto,nees:relat,dromey:Nature09}. An additional HOHG mechanism in the relativistic regime is coherent synchrotron emission (CSE)~\cite{daniel10,dromey:Nature12,dromey:NJP13}, which becomes important for thin targets in transmission geometry.

In case of the gas harmonics the spectrum exhibits a plateau in a certain range of higher orders before it finally rapidly decreases at some ``cut-off'' frequency~\cite{L'Huillier:hhg}. With the help of a spectral filter one can cut  out a range of harmonics with comparable amplitudes and get a single attosecond pulse~\cite{gouliemakis:single,corkum:subfemto}. Using only few harmonics~\cite{L’Huillier:photoionisation} allows carrying out experiments on the attosecond scale with trains of attosecond pulses still maintaining a certain spectral resolution.

The situation is different for the surface HOHG. The distribution of the harmonic intensity decreases rather steeply with the harmonic number $n$. Only deep in the relativistic regime a simple power law for the harmonic spectrum is obtained, e. g. $I(\omega_n)\propto n^{-8/3}$ for a normal incidence of the pump~\cite{baeva:theory,tar:JPB}, or slower for the oblique incidence~\cite{tar:JPB} or CSE regime~\cite{daniel10}. As a result, a spectrum can have a relatively small effective width and needs proper shaping in order to support attosecond pulses~\cite{tar:JPB}.

Moreover, for attosecond pulse production the harmonics have to be phase locked. A nonlinear harmonic spectral phase $\phi_{nl}(\omega)$ leads to a frequency variation (chirp) and broadening of the pulses in the time domain. The analysis of attosecond pulse trains in the case of gas harmonics demonstrated two different contributions to the nonlinear phase~\cite{varju:chirp}. A ``slow'' $\phi_{nl}(\omega)$ is responsible for the ``atto chirp'' and influences strongly the duration of the pulses in the train. A ``rapid'' phase change within the bandwidth of a single harmonic corresponds to the ``harmonic chirp'', which according to~\cite{varju:chirp} leads to shifts of the pulses in the attosecond pulse train. The same behavior is expected also for surface harmonics~\cite{atto:nomura}.

Simulations show that attosecond pulses can be “synthesized” using HOHG from surfaces by spectral filtering of the reflected waveform~\cite{rykovanov:single,krausz:ivanov,tsakiris:atto,baeva:theory,tar:JPB}. Methods for single attosecond pulse generation were discussed in~\cite{tsakiris:atto,baeva:relcont,naumova:atto,wheeler:nphot12}. For the first time the intensity autocorrelation function was measured in~\cite{atto:nomura,atto}. However, an explicit experimental evidence of reaching attosecond pulse durations remains a challenge.

For high contrast attosecond pulses spatial coherence of the harmonic beam is important. The influence of the surface roughness of the targets on the coherence has been investigated in~\cite{dromey:Nature09}. Fortunately, the manufacturing tolerances for roughness, which are typically 5~$nm$ and $<0.5~nm$ for high quality optics, allow high quality beams for up to the 40th to 400th harmonic of the Ti:Sa laser, respectively. More critical is the spatial coherence of the pump beam, which determines the effective harmonic source size. Moreover, for high-intensity laser-solid interactions pulses of very high contrast are required. A prepulse or a slowly rising leading front of the pump pulse lead to premature ionization of the target and destruction of the steep density profile before the arrival of the high intensity pulse maximum. This destruction accompanied by development of instabilities~\cite{tar:32harm} can spoil the harmonic spatial coherence~\cite{zhang:coh}. As shown in~\cite{tar:PRL07,tar:pra00,dromey:Nature09,hemmers:coh}, using high contrast femtosecond pulses allows harmonic emission with a divergence comparable with the diffraction limit. This indicates a high degree of spatial coherence. A corresponding direct measurement is still missing.

The attosecond pulse generation is usually associated with XUV carrier frequencies~\cite{tsakiris:atto,krausz:ivanov,Schultze:photo}, which is in fact not necessary. Synthesizing of attosecond pulses at considerably lower frequencies was demonstrated in~\cite{Wirth:SynthTrans}. The pulses were produced by proper adjustment of spectral amplitudes and phases of few spectral components of a supercontinuum, which was generated by femtosecond pulses. Such ``low carrier frequency'' attosecond pulses allowed tracking the nonlinear response of bound electrons in Kr atoms~\cite{Hassan:atto}. A wider spectral range from NIR to VUV can be covered by the surface HOHG in the CWE regime. Due to the resonant enhancement relatively high conversion efficiencies can be already achieved at non-relativistic intensities~\cite{tar:PRL07}.

Modifying the pump field has been shown to be a promising method of controlling the HOHG in gases. Two-pulse, two-color pump consisting of the driving wave and its second harmonic have been used for i) enhancement of the harmonic generation, ii) characterisation of the   emission on an attosecond scale, and iii) changing the period of the attosecond pulse trains (see~\cite{twopulse:Paulus,twopulse:dudovich,twopulse:kim,twopulse:Dahlstroem} and references therein). The corresponding intensity control in the case of the surface harmonics has been investigated in~\cite{tar:JPB,Yeung:NPhot,edwards:platonenko,twopulse:Salehi}.

In this paper we demonstrate shaping of harmonic spectra using single- and two-pulse pumping. The spectra consist of three harmonics either of lower (2nd to 4th) or of higher (12th to 14th) orders. The autocorrelation function of the total harmonic field is measured in the focused harmonic beam. For overlapping harmonics it perfectly corresponds to the measured spectrum. Furthermore, a high degree of harmonic coherence is demonstrated.

\begin{figure}[t]
\centering
\includegraphics[width=0.84\linewidth]{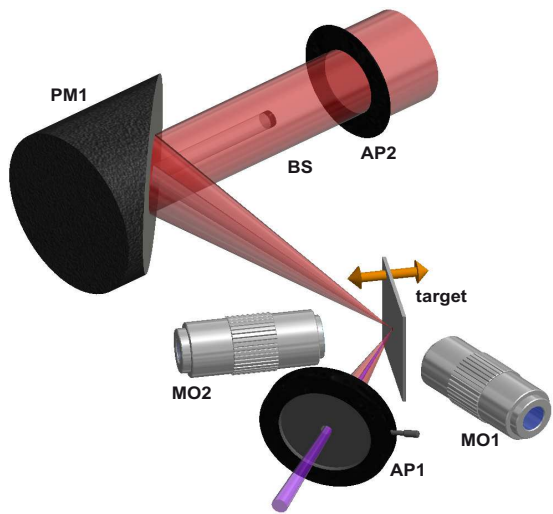}
\caption{\label{Fig1} (color online) Pump beam is focused onto the target with the help of the parabolic mirror PM1. The aperture AP1 together with the beam stop BS blocks the driver radiation~\cite{mirror:apperture}. The aperture AP2 with the diameter of 25~$mm$ cuts out the central part of the pump beam. Long working distance microscope objectives MO1 and MO2 are used for monitoring the energy distribution in the focal plane of PM1 and the target positioning (auto focus), respectively.}
\end{figure}
\section{\label{single}
Single pulse experiments}
The experimental setup is shown in Fig.~\ref{Fig1}. We used a Ti:Sapphire laser producing pulses of 45~$fs$ duration at  $\lambda$ = 800~$nm$ with a 10~$Hz$ repetition rate~\cite{tar:PRL07}. The pulses had a contrast ratio of $10^7$ at 1~$ps$ from the pulse maximum~\cite{tar:PRL07}. The wave front distortions of the laser beam were corrected with the help of an adaptive mirror. The laser beam was focused onto the target by the parabolic mirror PM1 with an effective focal length of about 75~$mm$. The focal spot distribution was monitored using the microscope objective MO1. The full width at half maximum (FWHM) was 2.5~$\mu m$ with about 30\% of the total pulse energy within the FWHM. At the maximum pulse energy on the target of 20~$mJ$ and the angle of incidence of \ang{40} the peak intensity reached $3\times10^{18}$~$W/cm^2$. The beam stop BS was imaged by the PM1 onto the aperture AP1 (diameter about 10~$mm$) in order to block the driver radiation~\cite{vdl:jenke,mirror:apperture}. The target was raster scanned in order to provide a fresh surface for each laser pulse. A computer controlled auto focus system kept the target surface in the focal plane of the parabolic mirror with an accuracy of 4~$\mu m$. The experiments were carried out in a vacuum chamber with a residual pressure of $10^{-3}$~$mbar$.
\begin{figure}[t]
\centering
\includegraphics[width=0.90\linewidth]{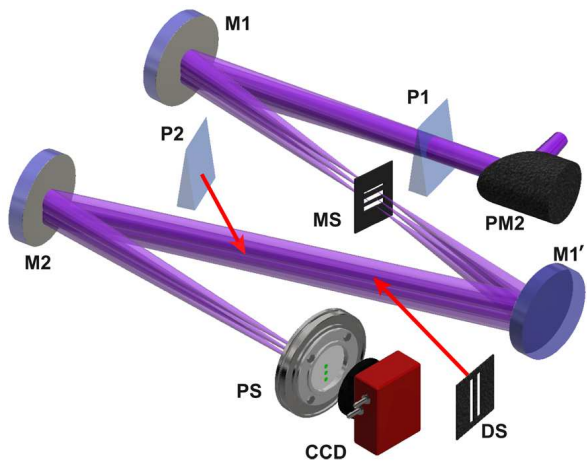}
\caption{\label{Fig2} (color online) LiF spectrometer: PM2, M1, M1$^\prime$~\cite{Note1}, and M2 are a parabolic mirror and concave spherical mirrors respectively. P1 and P2 are LiF prisms. Insertion of P2 compensates for the dispersion of P1. The mask MS placed in the focal plane of the mirror M1 adjusts the harmonic intensities. PS is a phosphor screen. The double slit DS is used to measure the harmonic temporal and spatial coherence.}
\end{figure}

The lower order harmonics were generated at a pump intensity of about $5\times10^{17}$~$W/cm^2$, which corresponds to the CWE regime of the HOHG~\cite{tar:PRL07}. Glass targets were used. The beam after AP1 (Fig.~\ref{Fig1}) was collimated with the parabolic mirror PM2 (Fig.~\ref{Fig2}) with an effective focal length of 170~$mm$ and spectrally dispersed with the help of the LiF prism P1 with an apex angle of \ang{10}. The intensity of individual harmonics could be manipulated with the help of the mask MS in the focal plane of the spherical mirror M1 ($F_{M1}$~=~500~$mm$), where the harmonic beams are spatially separated. The mask had three slits with effective widths of about 30~$\mu m$, 500~$\mu m$, and 500~$\mu m$ for the 2nd, 3rd, and 4th harmonic respectively. In addition the first two slits were covered with a 1~$mm$ thick UG11 filter in order to additionally attenuate the 2nd and the 3rd harmonic. The spherical mirrors M1$^\prime$ ($F_{M1^\prime}$~=~500~$mm$) and M2 ($F_{M2}$~=~750~$mm$) recollimated the harmonic beams and focused them onto the phosphor screen PS~\footnote{In Fig.~\ref{Fig2} an ``unfolded'' experimental schematic is shown for simplicity. Experimentally a folding plane mirror was placed in the focal plane of the mirror M1. The beam was reflected with a horizontal angular shift back on the mirror M1, which played the role of the mirror M1$^\prime$. The mask MS was placed close to the folding mirror.}. The screen had a nearly uniform spectral conversion efficiency for the harmonics of interest \cite{Phosphor}. The intensity distribution on the PS was observed with the help of a charge coupled device (CCD) camera.
\begin{figure}[t]
\centering
\includegraphics[width=1.0\linewidth]{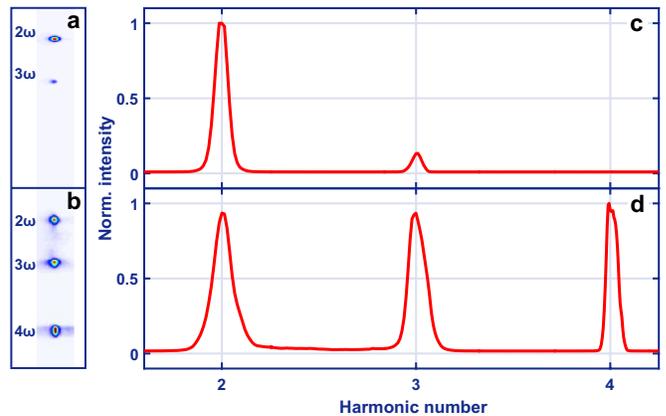}
\caption{\label{Fig3} (color online) CCD images (a,b) and the corresponding harmonic spectra (c,d). (a) and (c) show the initial spectrum, (b) and (d) the spectrum corrected by the mask MS.}
\end{figure}

Figure~\ref{Fig3} shows the CCD images (left) and the corresponding harmonic spectra (right) from the 2nd to the 4th order obtained with the help of the LiF spectrometer. The spectrometer was calibrated using both a standard mercury lamp and a hollow-cathode lamp~\cite{holcatlamp}. The upper and the lower panels were recorded without and with the mask MS respectively. The observed harmonic spectral widths are determined mostly by the spectrometer resolution. As expected, without the mask the harmonic intensity falls down rapidly with the harmonic order. Introducing the mask MS in the focal plane of the mirror M1 (see Fig.~\ref{Fig2}) allows to match the harmonic intensities.

In order to observe the harmonic coherence a double slit DS was placed into the beam after M1$^\prime$ (Fig.~\ref{Fig2}) The DS slits were oriented perpendicular to those of  MS. The slit widths $d$ and distance $R$ between them were 30~$\mu m$ and 300~$\mu m$, respectively. The corresponding interference pattern can be seen in Fig.~\ref{Fig4}a. Due to the angular dispersion introduced by the prism~P1 the pattern is spectrally resolved. The profiles of the individual harmonic patterns along the $x$-axis can be described by~\cite{BornWolf}:
\begin{figure}[t]
\centering
\includegraphics[width=1.0\linewidth]{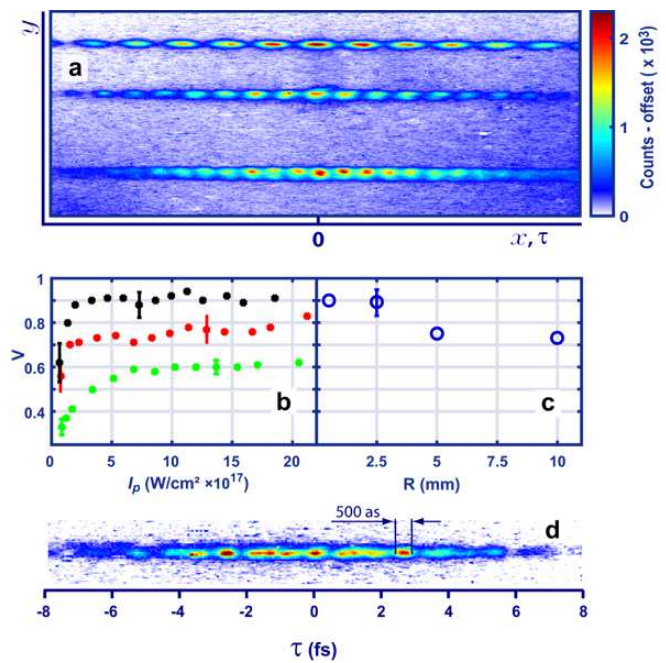}
\caption{\label{Fig4} (color online) (a) Spectrally resolved harmonic interference patterns. Top to bottom: 2nd, 3rd, and 4th harmonic patterns respectively. (b), (c) visibility V of the harmonic interference pattern as functions of $I_p$ (b) and the distance $R$ between the slits (c). \mbox{(\textbullet)} and \mbox{(\redbullet)} correspond to the 2nd harmonic with $R$ = 2.5~$mm$ and 5~$mm$, respectively, \mbox{(\textcolor{green}{\textbf{\textbullet}})} represents the visibility of the 3rd harmonic with $R$=0.5~$mm$. \mbox{(\textcolor{blue}{\large{\textbf{\textopenbullet}}})} corresponds to the 2nd harmonic at the pump intensity of $3\times10^{17}~W/cm^2$. (d) shows the interference pattern of the overlapping 2nd, 3rd, and 4th harmonic.}
\end{figure}
\begin{equation}
\label{eq1} I_I(x)=I^{(1)}(x)+I^{(2)}(x)+2\sqrt{I^{(1)}(x)}\sqrt{I^{(2)}(x)}Re[\gamma(\tau)],
\end{equation}
\noindent where $I_I$ is the time averaged intensity in the pattern, $I^{(i)}$ stands for an averaged intensity of a partial beam passing through a slit $i$, and $\gamma$ is the normalized field autocorrelation function (ACF):
\begin{equation}
\label{eq2} \gamma(\tau)=\left\langle E^{(1)}(x,t+\tau)E^{(2)}(x,t)\right\rangle/\sqrt{I^{(1)}(x)}\sqrt{I^{(2)}(x)},
\end{equation}

\noindent where $E^{(i)}$ are the electric fields of the partial waves and $\tau$ is the relative delay between them. Note that in case of the ultrashort pulses the time averaging indicated above with the sharp brackets is actually done over the entire pulse length. The delay $\tau$ originates from an angle $\theta$=$R/F_{M2}$ between the beams. It is related to the position on the $x$-axis (Fig.~\ref{Fig4}a) as $\tau$ = $xR/(cF_{M2})$, where c is the speed of light. In the following we consider $E^{(1)}(x,t)$ = $E^{(2)}(x,t)$ = $E(x,t)$ and $I^{(1)}$ = $I^{(2)}$ = $I$, which corresponds to the slit centered in the harmonic beam. In this case one has:
\begin{equation}
\label{eq3} I_I(x)=2I(x)(1+Re[\gamma(\tau)]),
\end{equation}
\noindent where $I(x) \propto \textrm{sinc}\left[\pi (d/\lambda) (x/F_{M2})\right]$ is determined by the diffraction on a single slit and the relation between $x$ and $\tau$ is given above. The width $\delta$ of the main maximum of $I(x)$ expressed in the number of periods of the corresponding interference pattern was $\delta$ = $R/d$ = $10$.

The spectra of the individual harmonics are relatively narrow. In this case it is natural to present $E(x,t)$ as $E(x,t)$ = $\mathcal{E}(x,t)\textrm{exp}(i\Omega t)+c.c.$, where $\mathcal{E}(t)$ is a slowly varying amplitude and $\Omega$ is the harmonic frequency. Assuming for simplicity that $\mathcal{E}(x,t)$ is real, one can rewrite (\ref{eq3}) as
\begin{equation}
\label{eq4} I_I(x)=2I(x)[1+\beta(\tau)\cos(\Omega\tau)],
\end{equation}
\noindent where $\beta(\tau)$ = $\left|\gamma(\tau)\right|$ is the normalized ACF of the field amplitude $\mathcal{E}$~\cite{akhmanov:opt}.

Figures~\ref{Fig4}b and \ref{Fig4}c represent the visibility~$V(x\approx0)$ = $(I_I^{max}-I_I^{min})/(I_I^{max}+I_I^{min})$ of the interference pattern as a function of the pump intensity and the distance between the slits. For equal intensities of the partial beams the visibility $V(R,x\approx0)$ = $\beta(R,\tau\approx0)$~\cite{BornWolf} characterizes the spatial coherence of the harmonic beam. Judging by Fig.~\ref{Fig4}b the spatial coherence stays high in a wide range of pump intensities and decreases for $I_p < 2\times10^{17}$~$W/cm^2$. The drop of $V$ at low $I_p$ can be attributed to the presence of broad-band plasma emission, which becomes comparable with the harmonic signal for low pump intensities. It is interesting to note that in gas harmonics the situation is different. There the coherence decreases with increasing pump intensity~\cite{ditmire:coh}. It can also be seen (Fig.~\ref{Fig4}c) that the coherence radius stays bigger than the beam diameter ($\approx$ 10~$mm$) determined by the aperture AP1~\footnote{According to the van Cittert-Zernike theorem~\cite{BornWolf} the harmonic source size $d\approx\lambda_nl/(\pi r_c)$, where $\lambda_n$ is a harmonic wavelength and $l$ the distance from the target. For $\lambda_n$ = 400~$nm$, $l$ = 170~$mm$, and $r_c > 10 mm$ one gets $d < 2.2 \mu m$, which corresponds very well to the pump focal spot.}.
\begin{figure}[t]
\centering
\includegraphics[width=0.95\linewidth]{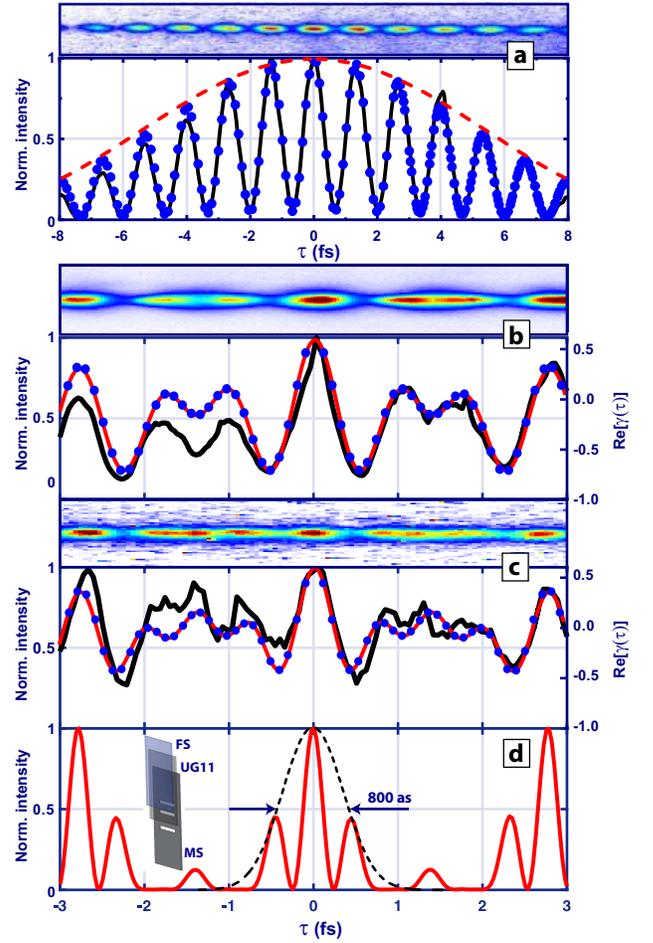}
\caption{\label{Fig5} (color online) Panels (a), (b), and (c) show the interference patterns and their profiles for: (a) 2nd, (b) overlapping 2nd and 3rd, (c) overlapping 2nd, 3rd, and 4th harmonic. The black curves correspond to experimental profiles. The dashed red curve in (a) depicts the normalized function $I(x)$ with $x$ = $\tau cF_{M2}/R$. The solid red lines and the blue circles represent profiles calculated with Re[$\gamma(\tau)$], which was obtained using the Wiener-Khinchin theorem and~(\ref{eq5}) respectively. The right axes for the profiles in (b) and (c) represent the Re[$\gamma(\tau)$]. Panel (d) shows a simulated pulse train corresponding to (c) with a compensated atto chirp. For the chirp compensation an additional fused silica platelet FS with the thickness of $\approx$700~$\mu m$ was placed in front of the mask MS and the filter UG11 (panel (d), inset).}
\end{figure}
\begin{figure*}[t!]
\centering
\includegraphics[width=0.75\linewidth]{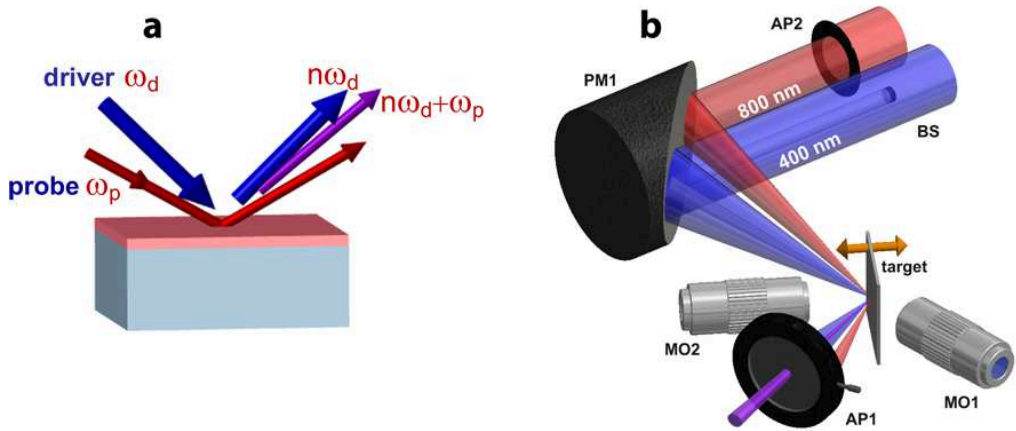}
\caption{\label{Fig6} (color online) (a) Two-pulse HOHG. The intensity of the driver at the frequency $\omega_d$ is much higher than that of the probe at the frequency $\omega_p$. In addition to the multiples of $\omega_d$ also hybrid-harmonics at the frequencies $n\omega_d+\omega_p$ are observed. (b) Experimental set-up. The parabolic mirror PM1 focuses the beams onto the target. The aperture AP1 and the beam stop BS are placed in the 400~$nm$ beam. The aperture AP2 controls the intensity of the 800~$nm$ beam on the target. The microscope objectives MO1 and MO2 i) monitor the focal energy distributions, ii) control auto focusing, and iii) monitor the beam overlapping.}
\end{figure*}

The ACF $\beta(\tau)/\beta(0)$ describes the harmonic temporal coherence. In Fig.~\ref{Fig5}a the profile of the 2nd harmonic pattern is compared to the one calculated according to~(\ref{eq4}) with the corresponding $I(x)$ and constant $\beta$ = $V$ = 0.9. The function $I(x)$ is also shown with $x$ = $\tau cF_{M2}/R$. One can see that the profile envelope is determined mostly by $I(x(\tau))$ with the half width at half maximum of about 6~$fs$. The correlation function stays approximately constant $\beta(\tau)\approx\ \beta(0)$ within this time, i.e. the correlation time $\tau_c\gg6$~$fs$~\footnote{one can roughly expect $\tau_c\approx\tau_p$, where $\tau_p$ is the pump pulse duration. A more detailed consideration of the relation $\tau_c/\tau_p$ can be found in~\cite{hemmers:coh}.}.

Now we turn to the ACF of the superposition of the harmonics (Fig.~\ref{Fig4}d). In order to observe it the prism P2 was placed between the mirrors M1$^\prime$ and M2 (see Fig.~\ref{Fig2}). The prism compensated for the angular dispersion in the beam. It can be seen that now the  interference pattern consists of sharp peaks with attosecond duration (correlation time $\tau_c<$ 500~$as$) and some broader background. The pattern period $T \approx\ $2.7~$fs$ corresponds to that of the pump wave. Unlike single harmonics it cannot be represented by~(\ref{eq4}) using a slowly varying $\beta(\tau)$, but expression~(\ref{eq3}) has to be used.

In general, $\gamma(\tau)$ is given by the Fourier transform of the spectrum (Wiener-Khinchin theorem \cite{BornWolf,akhmanov:opt}). Alternatively, an approximate expression for $\gamma(\tau)$ can be obtained directly in the time domain as follows. We assume that $I(x)$ and $\beta(\tau)$ are equal for all individual harmonics. This is reasonable within $\tau \lesssim T$ --- the time scale, which contains the features essential for the attosecond pulse production. Under this assumption we have:
\begin{equation}
\label{eq5} Re[\gamma(\tau)]=\beta(\tau)\frac{\sin(\frac{N}{2}\omega\tau)}{\sin(\frac{1}{2}\omega\tau)}\cos(\bar{\Omega}\tau),
\end{equation}

\noindent where $\omega$ is the fundamental frequency, $\bar{\Omega}$ is the central frequency of the harmonic spectrum, $N$ is the number of interfering harmonics, and $\beta(\tau)$ is the normalized ACF of an individual harmonic.

Figures~\ref{Fig5}b and \ref{Fig5}c  show the corresponding cutouts from the center of the interference patterns of two and three harmonics, respectively, with $\tau$ lying approximately within $\pm T$. Their profiles (black curves) are compared with the ones found with the help of~(\ref{eq3}). The red curves and the blue circles represent $Re[\gamma(\tau)]$ calculated from the  Fourier transform and~(\ref{eq5}) respectively. In the expression~(\ref{eq5}) we took $\beta$ = 0.6. The axes on the right hand side in Figs.~\ref{Fig5}b and \ref{Fig5}c correspond to $Re[\gamma(\tau)]$. We attribute the asymmetry in the experimental profiles to a slight misalignment of the overlapping harmonic beams with respect to each other.

One can see that the time domain picture (Fig.~\ref{Fig5}c) is in a good agreement with the spectral measurements (Fig.~\ref{Fig3}d) and can be well described by~(\ref{eq3}) and~(\ref{eq5}). This agreement suggests that the generated harmonics can indeed support attosecond pulses. Such a pulse is depicted in Fig.~\ref{Fig5}d. The pulse shape was simulated for the spectrum in Fig.~\ref{Fig3}d and the profile in Fig.~\ref{Fig5}c. The atto chirp (but not the harmonic chirp) was compensated by an additional fused silica platelet FS placed in front of the mask MS (covering the 30~$\mu$ slit, see inset in Fig.~\ref{Fig5}d) and shifting the prisms P1 and P2 in order to provide the proper amount of material dispersion in LiF. The harmonic propagation in the spectrometer and platelet was simulated using a ray-tracing code. The filter UG11 was also taken into account. The exact phase locking was achieved by slightly tilting the fused silica platelet.

\section{\label{two}
Two-pulse experiments}
The two-pulse-technique of the surface HOHG suggested in~\cite{tar:JPB} could be especially suitable for the attosecond pulse production using three harmonics as discussed above. The idea of this method is illustrated in Fig.~\ref{Fig6}a. A strong driver pulse with the frequency $\omega_d$ excites plasma oscillations and produces harmonics with the frequencies $n\omega_d$.  The weaker pulse with the frequency $\omega_p$, which we will call ``probe'', is mixed to the driver and gives rise to $n\omega_d+\omega_p$, which we will call hybrid-harmonics. The advantage of this technique is that both the amplitude and  the phase of the hybrid-harmonics can be adjusted independently from those of the driver-harmonics.

The driver-probe coupling and the amplification of the probe in the nonrelativistic regime occurs through a nonlinear excitation of plasma oscillations in a steep plasma-vacuum boundary by a component of the ponderomotive force given by $\textbf{F}\propto \nabla(\textbf{E}_d\textbf{E}_p)$, where $\textbf{E}_d$ and $\textbf{E}_p$ are the electric fields of the driver and the probe, respectively~\cite{tar:PRE10}. Generally speaking, the plasma oscillations corresponding to different harmonics are distributed within the plasma layer~\cite{tar:PRL07}, which may lead to the atto chirp of the reflected pulses (see~\cite{atto:nomura}, supplementary information).

Experimentally, the laser pulses at 800~$nm$ were frequency-doubled in a 0.8~$mm$ thick KDP crystal to produce high contrast 400~$nm$ pulses. The beams at 400~$nm$ and 800~$nm$ served as the driver and the probe, respectively. In this case the driver and the probe frequencies are $\omega_d$ = 2$\omega$ and $\omega_p$ = $\omega$, with the even driver-harmonics $2n\omega$ and odd hybrid-harmonics $(2n+1)\omega$.

Driver and probe were focused onto the target with the help of a parabolic mirror (Fig.~\ref{Fig6}b). The effective focal lengths of the mirror were 120~$mm$ and 75~$mm$ for the infrared and the blue beam, respectively. In the focal plane the beams FWHM were about 5~$\mu m$ at 800~$nm$ (AP2 fully open) and 1.5~$\mu m$ at 400~$nm$ with 40\% of the energy within the FWHM. The intensity of the driver was about $3\times10^{18}$~$W/cm^2$. The intensity of the probe was adjusted between $10^{16}-10^{17}$~$W/cm^2$ with the help of the aperture AP2. Mind that in spite of an angle between the driver and the probe, the hybrid and driver harmonics selected by the aperture AP1 are collinear with respect to each other.

\begin{figure}[t]
\centering
\includegraphics[width=0.9\linewidth]{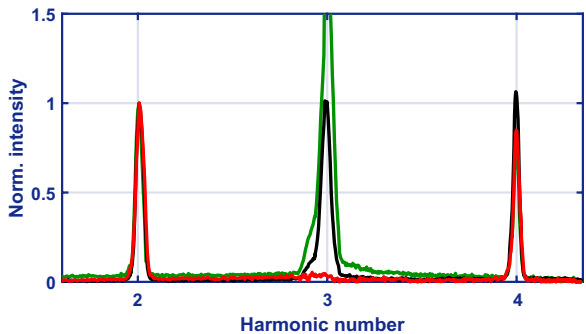}
\caption{\label{Fig7} (color online) Two-pulse harmonic spectra. The aperture AP2 (Fig.~\ref{Fig6}b) was closed~\mbox{(\textcolor{red}{\textbf{---}})} and had the diameters of 50~$mm$ (completely open)~\mbox{(\textcolor[rgb]{0.00,0.5,0.00}{\textbf{---}})}, and 6.8~$mm$~\mbox{(\textbf{---})}. The probe intensities were $10^{18}$~$W/cm^2$~\mbox{(\textcolor[rgb]{0.00,0.5,0.00}{\textbf{---}})} and about $10^{16}$~$W/cm^2$~\mbox{(\textbf{---})}.}
\end{figure}
\begin{figure*}[t]
\centering
\includegraphics[width=0.7\linewidth]{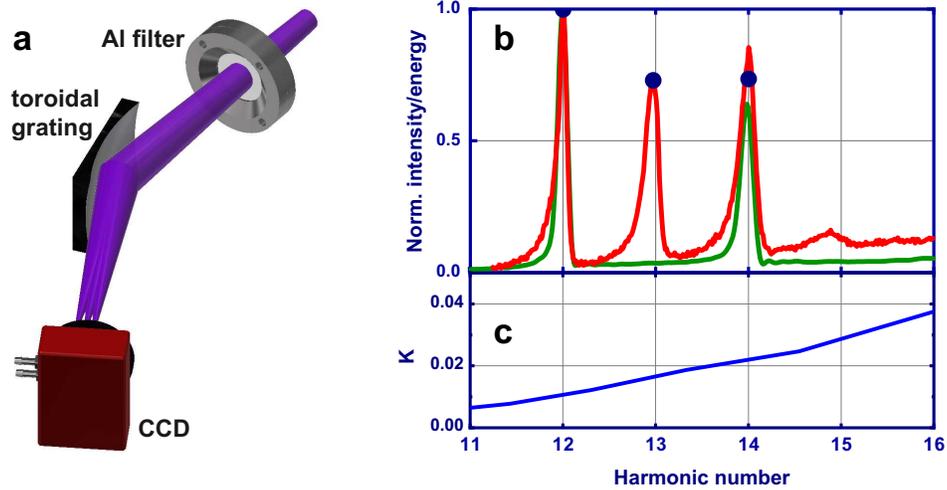}
\caption{\label{Fig8} (color online) Grating spectrometer (a): the toroidal grating images the plasma spot on the target spectrally resolved onto the CCD camera. The harmonic spectra~(b) as recorded by the CCD camera: aperture AP2 (Fig.~\ref{Fig6}b) closed \mbox{(\textcolor[rgb]{0.00,0.5,0.00}{\textbf{---}})} and open \mbox{(\textcolor{red}{\textbf{---}})}. Panel~(c) shows $K$=R$_{\textrm{G}} \times$ Eff$_{\textrm{CCD}}$ as a function of frequency. The circles \mbox{(\textcolor[rgb]{0.00,0.00,0.55}{\large{\textbullet}})} in (b) represent the energy of the individual harmonics integrated over the corresponding harmonic peak and corrected for $K$.}
\end{figure*}

Figure~\ref{Fig7} represents an example of the spectrum taken at different probe intensities using a glass target. It can be seen that a new (hybrid) component of the 3rd order appears in addition to the even ones. This is clear evidence of driver–-probe coupling leading to odd harmonics according to $\omega_{2n+1}$ = $2n\omega + \omega$. No mask or filters were used for the adjustment of the 3rd harmonic intensity. It was verified that no 3rd harmonic was observed when AP2 was blocked, or when a time delay between driver and probe was introduced. Another important feature is that although the intensity of the probe beam is much weaker than that of the driver, the intensity of the 3rd harmonic can be adjusted to be equal to the intensities of the 2nd and the 4th harmonic. For AP2 fully open the 3rd harmonic intensity is even higher, because the probe intensity becomes comparable to that of the driver.

The same behavior was observed with higher harmonics. The LiF spectrometers spectral range is limited by the 5th order harmonic due to absorption at higher orders. For this reason it was replaced by a spectrometer with a flat field toroidal grating as a dispersive element~(Fig.~\ref{Fig8}a). The grating spectrometer was also calibrated using the hollow-cathode lamp. A 170~$nm$ thick Al filter was used to block the fundamental and the lower harmonics up to the 11th order.

Polystyrene platelets were used as targets. Although glass targets demonstrated a somewhat higher harmonic efficiency, the HOHG spectrum from polystyrene has a cut-off after the 14th harmonic in the non-relativistic regime. This is lower compared to glass (20th harmonic)~\cite{tar:PRL07} and helped us to produce a three-harmonic spectrum. A spectrum measured with the driver alone is depicted in Fig.~\ref{Fig8}b (green curve). Due to the Al filter and the CWE cut-off only the 12th and the 14th harmonic with comparable intensities can be seen. The red curve in Fig.~\ref{Fig8}b represents the spectrum measured with both the driver and the probe. The probe intensity was adjusted so that the intensity of the additional hybrid-harmonic of the 13th order was close to those of the 12th and 14th harmonics.

The spectra of the individual harmonics are not resolved by the spectrometer. In order to compare the harmonic energies we integrate over the corresponding harmonic peak and take into account the grating reflectivity R$_{\textrm{G}}$ and the CCD efficiency Eff$_{\textrm{CCD}}$. The total response factor $K$=R$_{\textrm{G}} \times$ Eff$_{\textrm{CCD}}$ is plotted in Figure~\ref{Fig8}c as a function of frequency. The actual harmonic energies, corrected for $K$ are depicted in Fig.~\ref{Fig8}b by blue circles. One can see that the harmonic intensities can be well matched. As in the case of the lower harmonics it is possible to achieve a spectral shaping, which allows attosecond pulse generation.

\section{\label{conc}
Conclusions}
\label{sec:3}
In conclusion, we have demonstrated shaping of spectra consisting of three harmonics both for the lower and for the higher harmonic orders. For the lower harmonics the shaping was achieved using a spectral mask. The two-pulse-technique has shown to be a convenient way of controlling the amplitudes both for lower and higher orders. The harmonic emission possesses high spatial coherence, which stays constant in a wide range of pump intensities.

We have measured field autocorrelaton function of broadband harmonic radiation. To our knowledge, such correlation functions have never been recorded before. The recorded ACF exhibits spikes of attosecond duration, and its shape can be described by a simple analytical expression. A good correspondence between the frequency and time domain measurements indicates that we are not only producing the desired spectral shape in a spectrometer, but also overlap and focus the harmonics with the desired ratio of their intensities, which is necessary for attosecond pulse generation. In fact, the suggested measurement of the field autocorrelation provides a relatively simple proof of the combined focused harmonic field before a considerably more complicated intensity autocorrelation measurement is carried out.

The step to be done for the generation of corresponding attosecond pulses is the harmonic phase locking. In the single pulse case this can be achieved by adding a fused silica platelet to the optical path. Using two-pulse HOHG the phase adjustment can be reached simply by setting a proper time delay between driver and probe. Another advantage of the two-pulse-technique is that using appropriate spectral filtering it can be applied to harmonics ranging from NIR to XUV.
\begin{acknowledgments}
We would like to thank Prof. T.~Juestel for his help with the phosphor screen and the Deutsche Forschungsgemeinschaft (DFG, German Research Foundation) for its financial support through grants TA~292/4-1 and SFB 1242, project number 278162697.
\end{acknowledgments}
%\bibliography{Harmonics}
\end{document}